# Phase Mapping Of Remote Clocks Using Quantum Entanglement

M.S. Shahriar[1,2]

[1]*Dept. of Electrical and Computer Engineering, Northwestern University, Evanston, IL 60208*

[2]*Research Laboratory of Electronics, Massachusetts Institute of Technology, Cambridge, MA 02139*

**Abstract**

Recently, we have shown how the phase of an electromagnetic field can be determined by measuring the *population* of either of the two states of a two-level atomic system excited by this field, via the so-called Bloch-Siegert oscillation resulting from the interference between the co- and counter-rotating excitations. Here, we show how a degenerate entanglement, created without transmitting any timing signal, can be used to teleport this phase information. This phase-teleportation process may be applied to achieve absolute phase-mapping of remote clocks.

A recent proposal by Jozsa et al.[1] shows that if one has a singlet state of two atoms (consisting of entangled states of different but known energy), the intrinsic entanglement of the singlet can be used to create synchronized clocks which can be *simultaneously* started by making a measurement on one of the atoms. This is an important and interesting observation; however, all known methods for constructing a singlet state of two atoms using states of different energy require synchronized clocks to begin with. To put it differently, the challenge of quantum clock synchronization can be reduced to the task of determining the phase difference between two remote clocks, without transmitting a timing signal. This would enable phase-locking of clocks for applications in VLBI (very large base interferometry)[3] and for the global positioning system[4]. In this letter, we show a variation of this approach, whereby remote phase mapping can be achieved via teleporting the phase of a remote oscillator, via making use of the so-called Bloch-Siegert oscilllation, which results from an interference between the co- and counter-rotating parts of an excitation[5].

We assume that Alice and Bob each has an atom that has two degenerate ground states ($|1>$ and $|2>$), each of which is coupled to a higher energy state ($|3>$), as shown in figure 1. We assume the 1-3 and 2-3 transitions are magnetic dipolar, and orthogonal to each other, with a transition frequency $\omega$. For example, in the case of $^{87}$Rb, $|1>$ and $|2>$ correspond to $5^2P_{1/2}:|F=1,m_F=-1>$ and $5^2P_{1/2}:|F=1,m_F=1>$ magnetic sublevels, respectively, and $|3>$ corresponds to $5^2P_{1/2}:|F=2,m_F=0>$ magnetic sublevel. Left and right circularly polarized magnetic fields, perpendicular to the quantization axis, are used to excite the 1-3 and 2-3 transitions, respectively. The rubidium clock is typically stabilized with respect to the $5^2P_{1/2}:|F=1,m_F=0>$ to $5^2P_{1/2}:|F=2,m_F=0>$ transition, which is excited by a magnetic field parallel to the quantization axis[6]. We take $\omega$ to be the same as the clock frequency $\omega_c$.

We assume that Alice and Bob's fields at $\omega$ have the form $B_A=B_{ao}Cos(\omega t+\phi)$ and $B_B=B_{bo}Cos(\omega t+\chi)$, respectively. The origin of the time variable, $t$, is therefore arbitrary, and does not affect the quantity of interest: the phase difference, $\Omega\equiv(\phi-\chi)$. The clocks are assumed to be in phase if $\Omega=0$, so that if Bob determines that at some instant his magnetic field is maximum and positive in some direction $r_b$, then Alice will also find her magnetic field to be maximum and positive in some direction $r_a$ at the same instant. As long as Alice and Bob agree on this definition of phase-locking, and use the same definitions all the time, then $r_b$ and $r_a$ do not have to be the same. During the magnetic resonance excitations, the value of any dc magnetic field will be assumed to be vanishing. Symmetry then dictates that any physical observable will be independent of the choice of the quantization axis, as long as it is perpendicular to $r_a$ for Alice, and perpendicular to $r_b$ for Bob. Before we discuss our protocol, we summarize briefly a two-level interaction *without* RWA, and time reversal of an arbitrary evolution under this condition.

Consider, for example, the excitation of the $|1>_A \leftrightarrow |3>_A$ transition. In the dipole approximation, the Hamiltonian can be written as:

$$\hat{H} = \begin{bmatrix} 0 & g(t) \\ g(t) & \varepsilon \end{bmatrix} \qquad \ldots(1)$$

and the state vector is written as:

$$|\xi(t)\rangle = \begin{bmatrix} C_{1A} \\ C_{3A} \end{bmatrix} \qquad \ldots(2)$$

where $g(t) = -g_o[\exp(i\omega t+i\phi)+c.c.]/2$, and we assume that $\varepsilon=\omega$ corresponding to resonant excitation. Here, we have assumed that the polarization of Alice's field can be changed to excite either $|1>_A \leftrightarrow |3>_A$ or the $|2>_A \leftrightarrow |3>_A$ transition, in a mutually exclusive manner. We now perform a rotating wave transformation by operating on $|\xi(t)>$ with the unitary operator Q, given by:

$$\hat{Q} = \begin{bmatrix} 1 & 0 \\ 0 & \exp(i\omega t + i\phi) \end{bmatrix} \quad \ldots(3)$$

The Schroedinger equation then takes the form (setting $\hbar=1$):

$$\frac{\partial |\tilde{\xi}(t)>}{\partial t} = -i\tilde{H}(t)|\tilde{\xi}(t)> \quad \ldots(4)$$

where the effective Hamiltonian is given by:

$$\tilde{H} = \begin{bmatrix} 0 & \alpha(t) \\ \alpha^*(t) & 0 \end{bmatrix} \quad \ldots(5)$$

with $\alpha(t)= -g_o[\exp(-i2\omega t-i2\phi)+1]/2$, and the rotating frame state vector is:

$$|\tilde{\xi}(t)> \equiv \hat{Q}|\xi(t)> = \begin{bmatrix} \tilde{C}_{1A} \\ \tilde{C}_{3A} \end{bmatrix} \quad \ldots(6)$$

At this point, one may choose to make the rotating wave approximation (RWA), corresponding to dropping the fast oscillating term in $\alpha(t)$. As we will show, this corresponds to ignoring effects (such as the Bloch-Siegert shift) of the order of $(g_o/\omega)$, which can easily be observable if $g_o$ is large[7-10]. On the other hand, by choosing $g_o$ to be small enough, we can make the RWA for any value of $\omega$. We will make use of both regimes in our protocol. As such, we will now find the results without the RWA.

Given the periodic nature of the effective Hamiltonian, the general solution to eqn. 5 can be written as:

$$|\tilde{\xi}(t)> = \sum_{n=-\infty}^{\infty} |\xi_n\rangle \beta^n \quad \ldots(7)$$

where $\beta=\exp(-i2\omega t-i2\phi)$, and

$$|\xi_n\rangle \equiv \begin{bmatrix} a_n \\ b_n \end{bmatrix} \quad \ldots(8)$$

Inserting eqn. 7 in eqn. 4, and equating coefficients with same frequencies, we get, for all n:

$$\dot{a}_n = i2n\omega a_n + ig_o(b_n + b_{n-1})/2$$
$$\dot{b}_n = i2n\omega b_n + ig_o(a_n + a_{n+1})/2 \quad \ldots(9)$$

Figure 2 shows a pictorial representation of these equations. Here, the coupling between $a_o$ and $b_o$ is the conventional one present when the RWA is made. The coupling to additional levels results from virtual multiphoton processes in the absence of the RWA. The couplings to the nearest neighbors, $a_{\pm 1}$ and $b_{\pm 1}$ are detuned by an amount $2\omega$, and so on. To the lowest order in $(g_o/\omega)$, we can ignore terms with $|n|>1$, thus yielding a truncated set of six eqns.:

$$\overset{\circ}{a}_o = ig_o(b_o + b_{-1})/2 \qquad (10.1)$$

$$\overset{\circ}{b}_o = ig_o(a_o + a_1)/2 \qquad (10.2)$$

$$\overset{\circ}{a}_1 = i2\omega a_1 + ig_o(b_1 + b_o)/2 \qquad (10.3)$$

$$\overset{\circ}{b}_1 = i2\omega b_1 + ig_o a_1/2 \qquad (10.4)$$

$$\overset{\circ}{a}_{-1} = -i2\omega a_{-1} + ig_o b_{-1}/2 \qquad (10.5)$$

$$\overset{\circ}{b}_{-1} = -i2\omega b_{-1} + ig_o(a_{-1} + a_o)/2 \qquad (10.6)$$

In order to solve these equations, one may employ the method of adiabatic elimination valid to first order in $\sigma \equiv (g_o/4\omega)$. To see how this can be done, consider first the last two equations: 10.5 and 10.6. In order to simplify these two equations further, one needs to diagonalize the interaction between $a_{-1}$ and $b_{-1}$. To this end, we define $\mu_- \equiv (a_{-1} - b_{-1})$ and $\mu_+ \equiv (a_{-1} + b_{-1})$, which can be used to re-express these two equations in a symmetric form as:

$$\overset{\circ}{\mu}_- = -i(2\omega + g_o/2)\mu_- - ig_o a_o/2 \qquad (11)$$

$$\overset{\circ}{\mu}_+ = -i(2\omega - g_o/2)\mu_+ + ig_o a_o/2 \qquad (12)$$

Adiabatic following then yields (again, to lowest order in $\sigma$):

$$\mu_- \approx -\sigma a_o; \qquad \mu_+ \approx \sigma a_o \qquad (13)$$

which in turn yields:

$$a_{-1} \approx 0; \qquad b_{-1} \approx \sigma a_o \qquad (14)$$

In the same manner, we can solve eqns. 10.3 and 10.4, yielding:

$$a_1 \approx -\sigma b_o; \qquad b_1 \approx 0 \qquad (15)$$

Note that the amplitudes of $a_{-1}$ and $b_1$ are vanishing (each proportional to $\sigma^2$) to lowest order in $\sigma$, thereby justifying our truncation of the infinite set of relations in eqn. 9. Using eqns. 14 and 15 in eqns. 10.1 and 10.2, we get:

$$\overset{\circ}{a}_o = ig_o b_o/2 + i\Delta a_o/2 \qquad (16)$$

$$\overset{\circ}{b}_o = ig_o a_o/2 - i\Delta b_o/2 \qquad (17)$$

where $\Delta = g_o^2/4\omega$ is essentially the Bloch-Siegert shift. Eqns. 16 and 17 can be thought of as a two-level system excited by a field detuned by $\Delta$. With the initial condition of all the population in $|1>_A$ at $t=0$, the only non-vanishing (to lowest order in $\sigma$) terms in the solution of eqn. 9 are:

$$a_o(t) \approx Cos(g_o t/2); \quad b_o(t) \approx iSin(g_o t/2)$$
$$a_1(t) \approx -i\sigma Sin(g_o t/2); \quad b_{-1}(t) \approx \sigma Cos(g_o t/2) \qquad (18)$$

We have verified this solution via numerical integration of equation 10, as illustrated in figure 3. Inserting this solution in eqn. 7, and reversing the rotating wave transformation, we get the following expressions for the components of eqn. 2:

$$C_{1A}(t) = Cos(g_o t/2) - 2\sigma\Sigma \cdot Sin(g_o t/2)$$
$$C_{3A}(t) = ie^{-i(\omega t + \phi)}[Sin(g_o t/2) + 2\sigma\Sigma^* \cdot Cos(g_o t/2)] \qquad (19)$$

where we have defined $\Sigma \equiv (i/2)\exp[-i(2\omega t + 2\phi)]$. To lowest order in σ this solution is normalized at all times. Note that if Alice were to carry this excitation on an ensemble of atoms through for a π/2 pulse, and measure the population of the state $|1>_A$ immediately (at t=τ, the moment when the π/2 excitation ends), the result would be a signal given by $[1+2\sigma \sin(2\omega\tau+2\phi)]/2$, which contains information related to the amplitude and phase of her field.

Next, we consider the issue of exact time reversal of such an excitation. The Schroedinger eqn. (4) has the formal solution:

$$|\tilde{\xi}(t_2)> = \exp(-i\int_{t_1}^{t_2} \tilde{H}(t')dt')|\tilde{\xi}(t_1)> \qquad \ldots(20)$$

If the RWA is made, then $\tilde{H}$ is time independent. In that case, if one starts an evolution at $t_1$, proceed for *any* duration T, then reverses the sign of $\tilde{H}$ by shifting the phase of the magnetic field by π, and continues with the evolution for another duration T, then the system returns back to the starting state. Of course, this can be verified explicitly using the well known solution of Rabi flopping. Here, however, RWA is not made, so that $\tilde{H}$ depends on time. Therefore, the exact reversal can be achieved in this manner only if T=mπ/ω for any integer value of m. We have verified this conclusion via numerical integration as well (not shown). Parenthetically, note that while we have considered direct excitations of the two-level systems, all the results derived above apply equally to the case where an off-resonant Raman excitation is used to couple the two levels[11-14].

Returning to the task at hand, our protocol starts by using a scheme, developed earlier by us[2] (note that this scheme works for $^{87}$Rb, for the choice of |1> and |2> indicated above) to produce a degenerate entanglement of the form $|\psi>=(|1>_A|2>_B - |2>_A|1>_B)/\sqrt{2}$. Next, Alice attenuates her field so that the counter-rotating term in the Hamiltonian can be ignored (this assumption is not essential for our conclusion, but merely simplifies the algebra somewhat), and excites a π-pulse coupling $|2>_A$ to $|3>_A$, and then stops the excitation. Similarly, Bob uses a field, attenuated as above, to excite a π-pulse coupling $|2>_B$ to $|3>_B$, and then stops the excitation. Using digital communications over a classical channel, Alice and Bob wait until they both know that these excitations have been completed. The resulting state is then given by :

$$|\psi(t)> = [|1>_A|3>_B \exp(-i\omega t - i\chi) - |3>_A|1>_B \exp(-i\omega t - i\phi)]/\sqrt{2}. \qquad \ldots(21)$$

The next step is for Alice to make a measurement along the $|1>_A \leftrightarrow |3>_A$ transition. For this process, she chooses a much larger value of $g_o$, so that the RWA can not be made. The state she wants to measure is *the one that would result if one were to start from state $|1>_A$, and evolve the system for a π/2 pulse using this stronger $g_o$*:

$$|+\rangle_A \equiv \frac{1}{\sqrt{2}}\left[\{1-2\sigma\Sigma\}|1\rangle_A + ie^{-i(\omega t+\phi)}\{1+2\sigma\Sigma^*\}|3\rangle_A\right] \quad \ldots(22)$$

where we have made use of eqn. 19. The state orthogonal to $|+>_A$ results from a 3π/2 pulse:

$$|-\rangle_A \equiv \frac{1}{\sqrt{2}}\left[\{1+2\sigma\Sigma\}|1\rangle_A - ie^{-i(\omega t+\phi)}\{1-2\sigma\Sigma^*\}|3\rangle_A\right] \quad \ldots(23)$$

(Equivalently, the state of eqn. 23 results from a π/2 pulse excitation starting from $-i|3>_A$ ). To first order in σ, these two states are each normalized, and orthogonal to each other. As such, one can re-express the state of the two atoms in eqn. 21 as:

$$|\psi(t)\rangle = \frac{1}{\sqrt{2}}\left[|+\rangle_A|-\rangle_B - |-\rangle_A|+\rangle_B\right] \qquad \ldots(24)$$

where we have defined:

$$|+\rangle_B \equiv \frac{1}{\sqrt{2}}\left[\{1-2\sigma\Sigma\}|1\rangle_B + ie^{-i(\omega t+\chi)}\{1+2\sigma\Sigma^*\}|3\rangle_B\right] \qquad \ldots(25)$$

$$|-\rangle_B \equiv \frac{1}{\sqrt{2}}\left[\{1+2\sigma\Sigma\}|1\rangle_B - ie^{-i(\omega t+\chi)}\{1-2\sigma\Sigma^*\}|3\rangle_B\right] \qquad \ldots(26)$$

She can measure the state $|+\rangle_A$ by taking the following steps: (i) Shift the phase of the B-field by $\pi$, (ii) Fine tune the value of $g_o$ so that $g_o=\omega/2m$, for an integer value of m, (iii) apply the field for a duration of $T=\pi/2g_o$, and (iv) detect state $|1\rangle_A$. Note that the constraint on $g_o$ ensures that $T=m\pi/\omega$, which is necessary for time reversal to work in the absence of the RWA. Once Alice performs this measurement, the state for Bob collapses to $|-\rangle_B$, given in eqn. 26. Note that if $\sigma$ is neglected, then the measurement produces a $|-\rangle_B$ that contains no information about the phase of Alice's clock, which is analogous to the Jozsa protocol[1].

In the present case, $|-\rangle_B$ does contain information about the amplitude and the phase of Alice's clock signal. In order to decipher this, Bob measures his state $|1\rangle_B$. The probability of success is:

$$p_\phi \equiv |{}_B\langle 1|-\rangle_B|^2 = \frac{1}{2}[1+2\sigma Sin(2\phi)]. \qquad \ldots(27)$$

where we have kept terms only to the lowest order in $\sigma$. Of course, the value of $\phi$(mod $2\pi$), the phase difference, can not be determined from knowing Sin($2\phi$) alone. However, this whole process can be repeated after, for example, Alice shifts the phase of her B-field by $\pi/2$, so that Bob can determine the value of Cos($2\phi$). It is then possible to determine the value of $\phi$ (mod $2\pi$) unambiguously.

The overall process can be carried out in one of two ways. First, consider the situation where Alice and Bob starts with X pairs of atoms, and entangle each pair in the form of equation 24. Then, over a digital communication channel, Alice sends Bob a list of the M atoms she found in state $|1\rangle_A$ after performing her measurement process described above. Bob performs his measurement only on this subset of atoms. Suppose he finds L number of atoms in state $|1\rangle_B$. Then:

$$\eta \equiv (\frac{L}{M}-\frac{1}{2}) \rightarrow \sigma Sin(2\phi) \qquad , for\ large\ M \qquad \ldots(28)$$

Thus, the value of $\eta$ determined asymptotically for a large number of entangled pairs will reveal the value of Sin($2\phi$). Alternatively, if only a single pair of atoms is available, then the same result can be obtained by repeating the whole process X times, assuming that $\phi$ remains unchanged during the time needed for the process.

Note that what is determined by Bob is $\phi$, not $\Omega$. Thus, it is not possible to measure the absolute phase difference in this manner. In fact, one must transmit a timing signal in order to determine $\Omega$. This process is potentially hampered by the presence of undetermined fluctuations in the intervening pathlength. We have proven that in general quantum entanglement offers no advantage over a classical approach in determining $\Omega$ in the presence of such an undetermined source of noise.[15] However, one could use this approach of phase teleportation in order to achieve frequency locking of two remote oscillators. This is described in detail in reference 16, which essentially represents an appended version of this paper.

To summarize, previously we have shown how the phase of an electromagnetic field can be determined by measuring the population of either of the two states of a two-level atomic system excited by this field, via the so-called Bloch-Siegert oscillation resulting from the interference between the co- and counter-rotating excitations. Here, we show how a degenerate entanglement, created without transmitting any timing signal, can be used to teleport this phase information, thus making it possible to perform remote phase-mapping. The extension of this work to remote frequency-locking for clock synchronization is described in reference 16.

This work was supported by DARPA grant # F30602-01-2-0546 under the QUIST program, ARO grant # DAAD19-001-0177 under the MURI program, and NRO grant # NRO-000-00-C-0158.

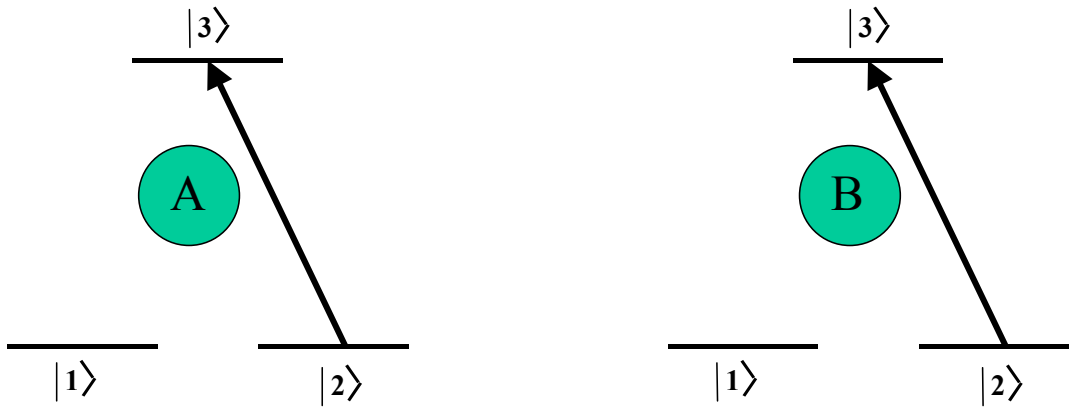

*Figure 1. Schematic illustration of the basic protocol for phase locking two remote clocks, one with Alice (A), and the other with Bob (B), without transmitting a clock signal directly. The model energy levels can be realized, for example, using the metastable hyperfine Zeeman sublevels of $^{87}$Rb atoms, as detailed in the text.*

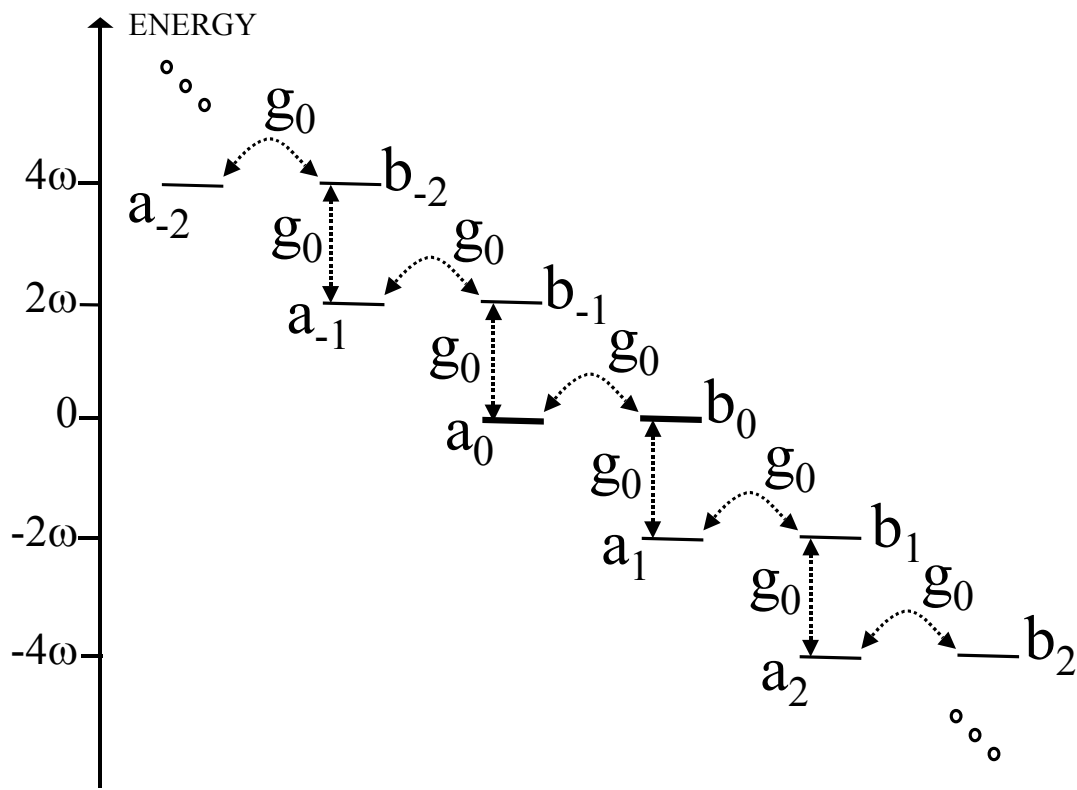

*Figure 2. Schematic illustration of the multiple orders of interaction when the rotating wave approximation is not made. The strengths of the first higher order interaction, for example, is weaker than the zeroth order interaction by the ratio of the Rabi frequency, $g_o$, and the effective detuning, $2\omega$.*

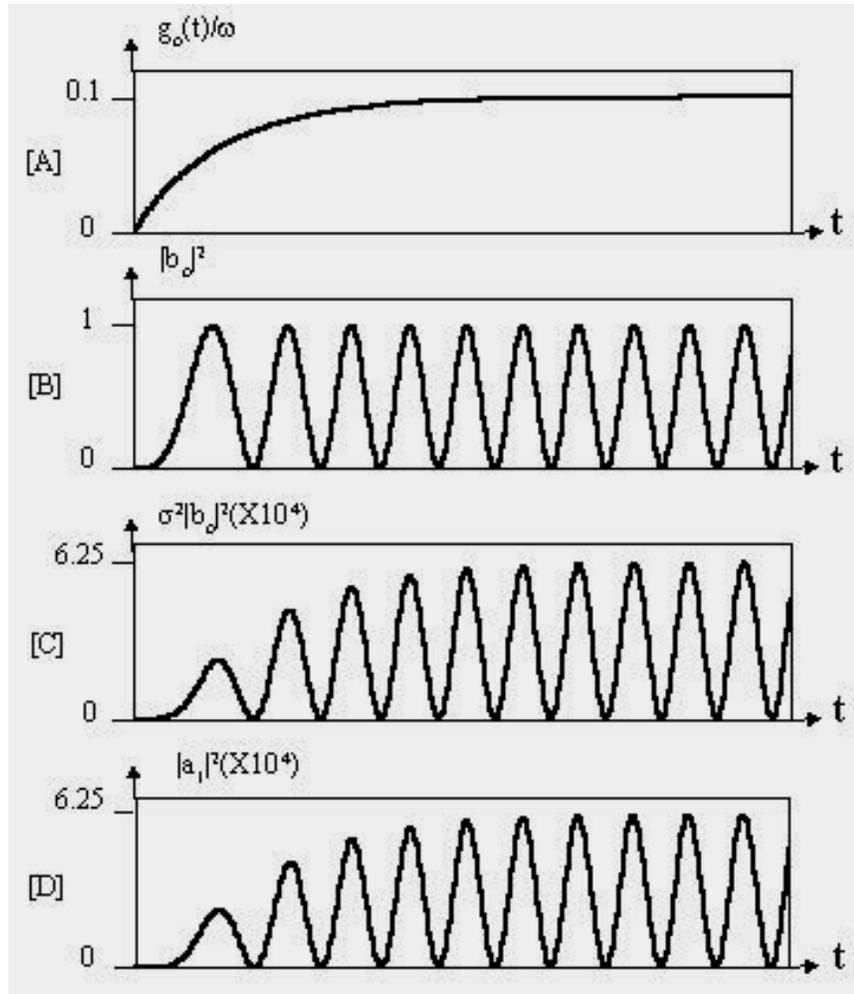

*Figure 3. Sample results from numerical integration of eqn. 10, confirming the result obtained analytically via adiabatic following: [A] The rabi frequency $g_o$ as a function of time, normalized to the constant transition frequency, $\omega$. The growth of $g_o$ is slow enough to allow for the adiabaticity condition to remain valid. [B] The corresponding evolution of $|b_o|^2$. [C] The value of $\sigma^2|b_o|^2$, which, according to the analytic solution is expected to be the value of $|a_1|^2$. [D] The value of $|a_1|^2$, which is virtually identical to $\sigma^2|b_o|^2$, thus confirming the analytic solution. The other components (e.g. $|b_{-1}|^2$) not shown here also agree with the analytical solution.*